\documentstyle[preprint,aps,epsf,floats]{revtex}

\begin{document}

\setlength\baselineskip{20pt}

\preprint{\tighten\vbox{\hbox{CALT-68-2229}\hbox{hep-ph/9907218}}}

\title{Bulk Fields in the Randall-Sundrum Compactification Scenario}

\author{Walter D. Goldberger\footnote{walter@theory.caltech.edu} and Mark B. Wise\footnote{wise@theory.caltech.edu}}
\address{\tighten California Institute of Technology, Pasadena, CA 91125}

\maketitle

{\tighten
\begin{abstract}
Recently, Randall and Sundrum proposed a solution to the hierarchy problem where the background spacetime is five dimensional.  There are two 3-branes, and the mass scale for fields that propagate on one of the 3-branes is exponentially suppressed relative to the fundamental scale of theory, which is taken to be the Planck mass, $M_{\hbox{\tiny{Pl}}}$.  In this letter we show that bulk fields with a five dimensional mass term of order $M_{\hbox{\tiny{Pl}}}$ have, after integrating over the extra dimension, modes with four-dimensional masses that are exponentially suppressed as well.  This opens the possibility that in this scenario the Standard Model matter fields may correspond to degrees of freedom that are not confined to a 3-brane.
\end{abstract}}
\vspace{0.7in}
\narrowtext

\newpage

It has been recently proposed that the existence of additional compact spatial dimensions could account for the large hierarchy between the Electroweak scale and the Planck scale~\cite{xdim}.  In these higher dimensional models, spacetime is usually taken to be the product of a four-dimensional spacetime and a compact $n$-manifold.  While gravity can propagate freely through the extra dimensions, Standard Model fields are confined to the four-dimensional spacetime.  Observers in this three-dimensional wall (a ``3-brane'') will then measure an effective Planck scale $M_{\hbox{\tiny{Pl}}}^2=M^{n+2} V_n,$ where $V_n$ is the volume of the compact space.  If $V_n$ is large enough, the fundamental Planck scale $M$ can be on the order of a TeV, removing the hierarchy between the weak and Planck scales.  Unfortunately, the presence of large extra dimensions does not necessarily provide a completely satisfactory resolution of the hierarchy problem, which resurfaces in the form of large ratio between $M$ and the compactification scale $\mu_c=V_n^{-1/n}.$ 

Randall and Sundrum~\cite{RS1} have proposed a higher dimensional scenario that is based on a non-factorizable geometry, and which accounts for the ratio between the Planck scale and weak scales without the need to introduce a large hierarchy between $M$ and $\mu_c.$  Their model consists of a spacetime with a single  $S^1/Z_2$ orbifold extra dimension.  Three-branes with opposite tensions reside at the orbifold fixed points, and together with a finely tuned negative bulk cosmological constant serve as  sources for five-dimensional gravity\footnote {\tighten This setup is analogous to the Horava-Witten scenario~\cite{HW} which arises in $M$-theory.  For a discussion of how the scenario described above Eq.~(\ref{eq:metric}) may arise from string theory compactifications, see~\cite{verlinde}.  Supergravity solutions which also exhibit exponential hierarchies are worked out in~\cite{supergrav}.  Extensions of Eq.~(\ref{eq:metric}) to curved 3-branes can be found in~\cite{curved}.  Cosmological implications are found in~\cite{cosmo}.}  .   The resulting spacetime metric contains a redshift factor which depends exponentially on the radius of the compactified dimension: 
\begin{equation}
\label{eq:metric}
ds^2 = e^{-2 k r_c |\phi|}\eta_{\mu\nu} dx^\mu dx^\nu - r_c^2 d\phi^2,
\end{equation}
where $x^\mu$ are Lorentz coordinates on the four-dimensional surfaces of constant $\phi$, and $-\pi\leq \phi\leq\pi$ (with $(x,\phi)$ and $(x,-\phi)$ identified, and the 3-branes located at $\phi=0,\pi).$  Here, $r_c$ sets the size of the extra dimension, and $k$ is taken to be on the order of the Planck scale.  

It is shown in~\cite{RS1} that the four-dimensional Planck scale is given by
\begin{equation}
M_{\hbox{\tiny Pl}}^2=\frac{M^3}{k}[1-e^{-2kr_c\pi}],
\end{equation}
so that $M_{\hbox{\tiny Pl}}$ is of order $M.$  Also, a field confined to the 3-brane at $\phi=\pi$ with mass parameter $m_0$ will have a physical mass given by $m=m_0 e^{-kr_c\pi}.$  Thus, if $k r_c$ is around 12 the weak scale is dynamically generated while all fundamental mass scales are on the order of the Planck scale, i.e. there is no very large hierarchy between $M$ and $\mu_c=1/r_c.$  Besides this, it is also found that on account of the exponential redshift factor in Eq.~(\ref{eq:metric}), Kaluza-Klein gravitational modes in this spacetime have TeV scale mass splittings and couplings~\cite{RS2}.  This is in sharp contrast to the Kaluza-Klein decomposition in product spacetimes, which for a large compactified dimensions, gives rise to a high number of light modes (with splittings of the order of the compactification scale) that are coupled only with gravitational strength.  The large number of light Kaluza-Klein modes has important phenomenological implications~\cite{particle}.  For example, astrophysical constraints imply that for two extra dimensions $M>30\hbox{ TeV}$~\cite{astro}.

In this letter, we extend these results by carrying out the Kaluza-Klein decomposition of a non-gravitational scalar bulk field\footnote{\tighten Other aspects of field configurations in the bulk have been studied in~\cite{mersini}.} propagating in the spacetime corresponding to Eq.~(\ref{eq:metric}).  Naively, one would expect even the lightest Kaluza-Klein modes to have masses comparable to the mass of the bulk scalar and to have self-interactions set by the Planck scale.  Instead we find that the mass spectrum of the four-dimensional Kaluza-Klein modes is suppressed by a factor of $e^{-kr_c\pi}$ relative to the five-dimensional scalar mass.  We also find that the same exponential factor suppresses the scale that sets the non-renormalizable self-couplings of the light modes.  If the bulk field mass and self-couplings are set by the Planck scale and as above $kr_c$ is around 12, the low-lying Kaluza-Klein modes would be characterized by a scale which is on the order of a TeV and could therefore have significant phenomenological consequences:  Standard Model particles could be low-lying Kaluza-Klein excitations of bulk fields.

First consider a free scalar field in the bulk.  The action is 
\begin{equation}
S={1\over 2}\int d^4 x\int_{-\pi}^\pi d\phi \sqrt{G} \left(G^{AB}\partial_A \Phi \partial_B \Phi - m^2 \Phi^2\right),
\end{equation}
where $G_{MN}$ with $A,B=\mu,\phi$ is given by Eq.~(\ref{eq:metric}), and $m$ is of order $M.$  After an integration by parts, this can be written as
\begin{equation}
\label{eq:action}
S={1\over 2}\int d^4 x\int_{-\pi}^\pi r_c d\phi  \left(e^{-2\sigma(\phi)}\eta^{\mu\nu}\partial_\mu\Phi \partial_\nu \Phi + {1\over r_c^2}\Phi\partial_\phi\left(e^{-4\sigma(\phi)}\partial_\phi\Phi\right)-m^2 e^{-4\sigma(\phi)}\Phi^2\right),
\end{equation}
with $\sigma(\phi)=kr_c|\phi|$.  To perform the Kaluza-Klein decomposition, write $\Phi(x,\phi)$ as a sum over modes:
\begin{equation}
\Phi(x,\phi)=\sum_n \psi_n(x) \frac{y_n(\phi)}{\sqrt{r_c}}.
\end{equation}
If the $y_n(\phi)$ are chosen to satisfy
\begin{equation}
\label{eq:ortho}
\int_{-\pi}^\pi d\phi e^{-2\sigma(\phi)} y_n(\phi) y_m(\phi)=\delta_{nm}
\end{equation}
and
\begin{equation}
\label{eq:eigen}
-{1\over r_c^2}\frac{d}{d\phi}\left(e^{-4\sigma(\phi)}\frac{dy_n}{d\phi}\right)+m^2 e^{-4\sigma(\phi)}y_n=m_n^2 e^{-2\sigma(\phi)} y_n,
\end{equation}
then, Eq.~(\ref{eq:action}) simplifies to
\begin{equation}
S={1\over 2}\sum_n \int d^4 x \left[\eta^{\mu\nu}\partial_\mu \psi_n \partial_\nu \psi_n - m_n^2 \psi_n^2\right].
\end{equation}
As in usual Kaluza-Klein compactifications, the bulk field $\Phi(x,\phi)$ manifests itself to a four-dimensional observer as an infinite ``tower'' of scalars $\psi_n(x)$ with masses $m_n$ which we find by solving the above eigenvalue problem.  After changing variables to $z_n =m_n e^{\sigma(\phi)}/k$ and $f_n = e^{-2\sigma(\phi)}y_n,$ Eq.~(\ref{eq:eigen}) can be written as (for $\phi\neq 0, \pm\pi$)
\begin{equation}
z_n^2 \frac{d^2 f_n}{d z_n^2} + z_n \frac{d f_n}{dz_n} + \left[z_n^2 - \left(4+\frac{m^2}{k^2}\right)\right]f_n=0.
\end{equation}
The solutions of this equation are Bessel functions of order $\nu=\sqrt{4+{m^2\over k^2}}$.  We thus find
\begin{equation}
\label{eq:solution}
y_n(\phi)=\frac{e^{2\sigma(\phi)}}{N_n}\left[J_\nu \left({m_n\over k} e^{\sigma(\phi)}\right) + b_{n\nu} Y_\nu \left({m_n\over k} e^{\sigma(\phi)}\right)\right],
\end{equation}
where $N_n$ is a normalization factor.  The condition that the differential operator on the LHS of Eq.~(\ref{eq:eigen}) be self-adjoint forces the derivative of $y_n(\phi)$ to be continuous at the orbifold fixed points.  This gives two relations that can be used to solve for $m_n$ and $b_{n\nu}$, yielding
\begin{equation}
\label{eq:coefs}
b_{n\nu}=-\frac{2 J_\nu \left(\frac{m_n}{k}\right)+\frac{m_n}{k} J'_\nu \left(\frac{m_n}{k}\right)}{2 Y_\nu \left(\frac{m_n}{k}\right)+\frac{m_n}{k} Y'_\nu \left(\frac{m_n}{k}\right)},
\end{equation}
and
\begin{eqnarray}
\label{eq:zeros}
0 &=& [2 J_\nu (x_{n\nu}) + x_{n\nu} J'_\nu (x_{n\nu})] [2 Y_\nu (x_{n\nu} e^{-kr_c\pi}) + x_{n\nu} e^{-kr_c\pi} Y'_{\nu} (x_{n\nu}e^{-kr_c\pi})]\nonumber \\
& & {}-[2 Y_\nu (x_{n\nu}) + x_{n\nu} Y'_\nu (x_{n\nu})] [2 J_\nu (x_{n\nu} e^{-kr_c\pi}) + x_{n\nu} e^{-kr_c\pi} J'_\nu (x_{n\nu}e^{-kr_c\pi})],
\end{eqnarray}
where $x_{n\nu}=m_n e^{kr_c\pi}/k.$  For $e^{kr_c\pi}\gg 1$, this last condition simplifies to
\begin{equation}
2 J_\nu (x_{n\nu}) + x_{n\nu} J'_\nu (x_{n\nu})=0.
\end{equation}
\begin{figure}[t!]
\centerline{\epsfysize=180pt \epsfxsize=285pt \epsfbox{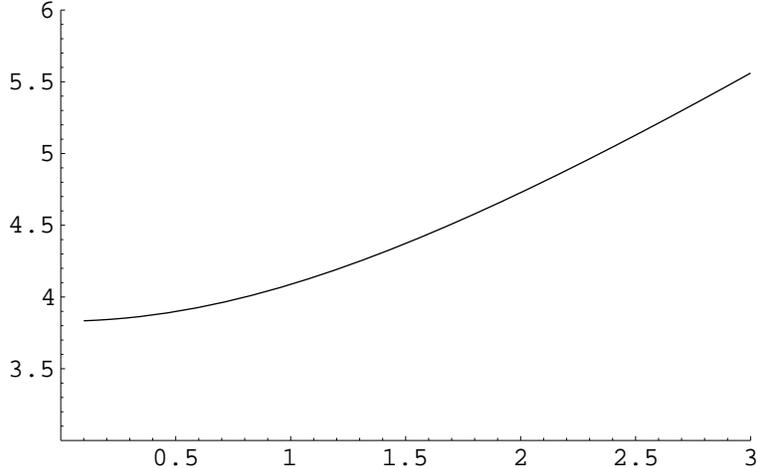}}
\caption{Plot of $x_{1\nu}$ versus  $m/k$ in the region where $m/k$ is order unity.}\label{fig}
\end{figure}
Fig.~\ref{fig} shows the lowest root of Eq.~(\ref{eq:zeros}), $x_{1\nu},$ as a function of $m/k$.  Because the lightest modes have $x_{n\nu}$ of order unity, we see that their masses are suppressed exponentially with respect to the scale $m$ appearing in Eq.~(\ref{eq:action}).  Since we take $m$ of order the Planck scale and $kr_c$ around 12 these light modes have masses in the TeV range.  The exponential suppression can be understood from Eq.~(\ref{eq:solution}):  the modes $y_n(\phi)$ are larger near the 3-brane at $\phi=\pi,$ and consequently it is more likely to find the Kaluza-Klein excitations in that region.  Therefore their masses behave in the same way as masses of fields confined to the brane at $\phi=\pi,$ which are characterized by the TeV scale.  

For the massless case, $m=0$, there is a mode with $y_1$ constant and $x_{12}=0$.  It can be obtained from Eq.~(\ref{eq:solution}) and Eq.~(\ref{eq:coefs}) by a limiting procedure.  When $x_{1\nu}$ is small Eq.~(\ref{eq:zeros}) yields
\begin{equation}
x_{1\nu}\simeq \frac{1}{\sqrt{2}}\left(\frac{m}{k}\right) e^{k r_c\pi}.
\end{equation}
Consequently $x_{1\nu}$ increases to near the minimum value shown in Fig.~\ref{fig} over an exponentially small region of $m/k.$  For the remainder of this paper we assume $m/k$ is of order unity.

For the low lying modes, the coefficient $b_{n\nu}$ is of order $e^{-2 \nu k r_c\pi}$ and we can safely ignore the $Y_\nu(z_n)$ term with respect to $J_\nu(z_n)$ when performing integrals involving the $y_n(\phi).$  Thus, to a good approximation
\begin{equation}
N_n\simeq \frac{e^{kr_c\pi}}{\sqrt{kr_c}} A_n,
\end{equation}
where
\begin{equation}
A_n=J_\nu(x_{n\nu})\sqrt{1+\frac{4-\nu^2}{x_{n\nu}^2}}.
\end{equation}

We now turn our attention to possible self-interactions of the bulk scalar.  From the four-dimensional point of view, these induce couplings between the Kaluza-Klein modes.  Here, we concentrate on the self-couplings of the light modes.  Just as in the case of the mass spectrum, we find that the exponential factor in Eq.~(\ref{eq:metric}) plays a crucial role in determining the effective scale of the couplings.  If the Planck scale sets the scale of the five-dimensional couplings, the low-lying Kaluza-Klein modes have TeV range self-interactions.  First consider a term in the action which is of the form:
\begin{equation}
S_{\hbox{\tiny{int}}}=\int d^4 x\int_{-\pi}^\pi d\phi\sqrt{G} \frac{\lambda}{M^{3m-5}}\Phi^{2m},  
\end{equation}
where $\lambda$ is of order unity.  Expanding in modes, the self-interactions of the light Kaluza-Klein states are given by 
\begin{equation}
\label{eq:int}
S_{\hbox{\tiny{int}}}=\int d^4 x\int_{-\pi}^\pi r_c d\phi  e^{-4\sigma(\phi)} \frac{\lambda}{M^{3m-5}}\psi_n^{2m} \left(\frac{y_n}{\sqrt{r_c}}\right)^{2m}.
\end{equation}
Thus, the effective four-dimensional coupling constants for the $\psi_n^{2m}$ interactions are
\begin{equation}
\lambda_{\hbox{\tiny{eff}}}=\frac{2\lambda}{(M r_c)^{m-1} M^{2m-4} (kr_c)}\int_0^{k r_c\pi} d\sigma e^{-4\sigma} y_n^{2m},
\end{equation}
which in the large $kr_c$ limit become
\begin{equation}
\lambda_{\hbox{\tiny{eff}}}\simeq 2\lambda\left(\frac{k}{M}\right)^{m-1} (M e^{-kr_c\pi})^{4-2m}\int_0^1 r^{4m-5} dr \left[\frac{J_\nu (x_{n\nu}r)}{A_n}\right]^{2m}.
\end{equation}
 Such couplings can also be induced by derivative self-interactions of the bulk field.  For example, the term 
\begin{equation}
S_{\hbox{\tiny{int}}}=\int d^4 x\int_{-\pi}^\pi d\phi \sqrt{G}\frac{\lambda}{M^{5m-5}} \left(G^{AB}\partial_A \Phi \partial_B\Phi\right)^m
\end{equation}
has a piece which contains only derivatives with respect to $\phi:$  
\begin{equation}
S_{\hbox{\tiny{int}}}=\int d^4 x\int_{-\pi}^\pi r_c d\phi  e^{-4\sigma(\phi)} \frac{\lambda}{M^{5m-5}} \psi_n^{2 m}\left(\frac{(\partial_\phi y_n)^2}{r_c^3}\right)^m.
\end{equation}
From the point of view of four-dimensional observers, this yields a $\psi_n^{2m}$ interaction with a coupling constant
\begin{equation}
\lambda_{\hbox{\tiny{eff}}}=\frac{2\lambda (kr_c)^{2m-1}}{(Mr_c)^{3m-1} M^{2m-4}}\int_0^{kr_c\pi} d\sigma e^{-4\sigma}\left(\frac{dy_n}{d\sigma}\right)^{2m}.
\end{equation}
For large $kr_c,$ this becomes
\begin{equation}
\lambda_{\hbox{\tiny{eff}}}\simeq 2\lambda\left(\frac{k}{M}\right)^{3m-1} (Me^{-kr_c\pi})^{4-2m}\int_0^1 r^{2m-5} dr \left[\frac{d}{dr}\left(r^2 \frac{J_\nu(x_{n\nu}r)}{A_n}\right)\right]^{2m}.
\end{equation}
In either case, we see that the the scale relevant to four-dimensional physics is not $M$, but rather $v=M e^{-kr_c\pi}.$  The Kaluza-Klein reduction has lead to an exponential enhancement of irrelevant couplings from Planck scale to only TeV scale suppression.

We have seen that in the Randall-Sundrum compactification scenario, bulk scalars have low lying Kaluza-Klein modes with four-dimensional masses of order the weak scale and four-dimensional non-renormalizable interactions suppressed by powers of the weak scale even though from a five-dimensional perspective their masses and interactions are characterized by the Planck scale.  This is similar to what occurs for the fields localized on the 3-brane at $\phi=\pi$.  It happens because the Kaluza-Klein eigenfunctions $y_n(\phi)$ are dominated by the region of $\phi$ near $\pi$.  Note that this is also similar to the Kaluza-Klein excitations of the graviton field~\cite{RS2}.  However, the zero mode (i.e. the four-dimensional graviton) eigenfunction is dominated by the region of $\phi$ near $0$, and from a four-dimensional perspective, gravitational interactions are still suppressed by the Planck scale.  

Results analogous to those found here for scalars also hold for vector fields.  Bulk fields can be Standard Model fields.  However, the construction of even semi-realistic models based on these ideas awaits a mechanism for generating a potential to stabilize the radius of the extra dimension, $r_c.$

We thank Petr Horava and Raman Sundrum for helpful discussions.  This work was supported in part by the Department of Energy under grant number DE-FG03-92-ER 40701.

\end{document}